\documentclass[%
 reprint,
 showpacs,preprintnumbers,
 amsmath,amssymb,
 aps,
 prb,
]{revtex4-1}
\usepackage{dcolumn}
\usepackage[dvipdfmx]{graphicx}
\usepackage{subfigure}
\usepackage{color}
\usepackage{mathrsfs}

\makeatletter
\def\btt#1{\texttt{\@backslashchar#1}}
\DeclareRobustCommand\bblash{\btt{\@backslashchar}} \makeatother

\begin{document}

\title{Anomalous Hall effect in $2H$-phase transition-metal dichalcogenide monolayers on ferromagnetic substrates}

\author{Tetsuro Habe and Mikito Koshino}
\affiliation{Department of Physics, Osaka University, Toyonaka, Osaka 560-0043, Japan}

\date{\today}

\begin{abstract}
We study the anomalous Hall effect in monolayers of transition-metal dichalcogenides
2H-$MX_2$ ($M$=Mo, W, and $X$=S, Se, Te) under a proximity effect of ferromagnetic substrate.
If a proximity-induced exchange field is introduced,
the spin-polarized energy bands in $K$ and $K'$ valleys are shifted in the opposite directions, 
and it causes the Hall effect by breaking time-reversal symmetry.
The induced Hall effect is the most prominent in the valence band which has a large intrinsic spin splitting.
Moreover, we find that tilting the magnetization from the perpendicular direction
gives rise to a sensitive change in the Hall conductivity only in the electron side,
and it is attributed to the mixing of the Berry phase by the in-plane field in the nearly degenerate conduction bands.
\end{abstract}


\maketitle
\section{Introduction}
Monolayers of transition-metal dichalcogenides (TMDCs) are atomically thin semiconductors with a direct energy gap at two symmetric points $K$ and $K'$ (called valleys) in the first Brillouin zone\cite{Helveg2000,Mak2010,Splendiani2010,Eknapakul2014}.
The low-energy electronic states can be described by the massive Dirac Hamiltonian, where the eigenstates have nontrivial Berry phase\cite{Haldane2004,Nagaosa2010,Xiao2012,Mak2014}.
The Berry phase leads to valley Hall effect in which the two valleys contribute to the opposite Hall currents\cite{Xiao2012}, while the net Hall conductivity exactly vanishes due to time-reversal symmetry.
The valley Hall effect in TMDC was experimentally detected by using the optical technique, where time-reversal symmetry is explicitly broken by the circular polarized light.\cite{Xiao2012,Zeng2012,Mak2012}.

In this paper, we theoretically investigate the anomalous Hall effect in the monolayer 2H-TMDC placed in close proximity with a ferromagnet.
In the intrinsic TMDC monolayer, the energy bands of $K$ and $K'$ valleys are spin-split in the opposite directions due to the spin-orbit interaction.\cite{Jin2013,Alidoust2014}
If a proximity-induced Zeeman field is introduced to the system, the band edges in the two valleys are shifted in the opposite directions in energy, and it leads to a static net Hall current by breaking the intervalley balance.
The Hall conductivity becomes the maximum near the valence band edge due to the large intrinsic spin split.
The effect in the conduction band is strongly depending on the transition-metal atom, where the Hall conductivity in Mo$X_2$ is similar to the valence band with the opposite sign and the effect in W$X_2$ is relatively weaker in the conduction band where the two spin states are nearly degenerate because still the difference in the effective mass causes a finite Hall conductivity.

The Hall effect is mainly caused by the magnetization component perpendicular to the layer, i.e., parallel to the spin splitting direction in the TMDC monolayer.
However, we also find that tilting the magnetization from the perpendicular direction gives rise to a sensitive change in the Hall conductivity in the electron side.
The effect is prominent particularly at the crossing point of two spin branches of the conduction band, where the Hall conductivity in the perpendicular magnetization nearly vanishes by only a few degree tilt.
The sensitive response to the magnetization direction is attributed to the mixing of the Berry phase in the two spin states by the in-plane field.

Generally, the anomalous Hall effect occurs in the presence of spin-orbit coupling and magnetization.\cite{Qi2008,Nagaosa2010}
The proximity-induced anomalous Hall effect was also studied for graphene on the ferromagnetic substrate theoretically \cite{Qiao2010,Qiao2014} and experimentally \cite{Wang2015},
where both the spin-orbit coupling and the magnetization are externally induced by the ferromagnetic substrate.
The anomalous Hall effect studied here relies on the intrinsic strong spin-orbit coupling and intrinsic Berry phase at two valleys, a characteristic property, in TMDC.

\section{Effective Hamiltonian}
We consider the electric states near the band edge in the monolayer 2H-TMDC, which can be described by the effective Hamiltonian for the relative wave vector $\boldsymbol{k}$ with respect to the valley points\cite{Xiao2012,Kormanyos2015},
\begin{align}
H_0=&v(\tau k_x\sigma_x+k_y\sigma_y)+\frac{\Delta}{2}\sigma_z\notag\\
&-\lambda_v\tau\frac{\sigma_z-1}{2}s_z-\lambda_c\tau\frac{\sigma_z+1}{2}s_z.\label{Hamiltonian}
\end{align}
The first and the second terms are the band Hamiltonian without the spin-orbit coupling, where the $v$ and $\Delta$ are the parameters for velocity and gap, respectively, $\tau=\pm1$  are the valley indexes for $K$ and $K'$, respectively, and $s_\mu$ is the Pauli matrix in the spin space. $\sigma_\mu$ is a Pauli matrix defined for two bases of the atomic orbitals $d_0$ and $d_{\pm2}$ for $\tau=\pm$, respectively, where $d_j$ is the $d$-orbital of the transition-metal atom with the orbital angular momentum $j$.
Third term in Eq. (\ref{Hamiltonian}) represents the spin-orbit coupling which is represented by a product of the spin angular momentum $s_z$ and the orbital angular momentum $\sigma_z-1$ with the coupling constant $\lambda_v$.
Forth term is also spin-orbit coupling associated with the spin-splitting in the conduction band, where the coupling is attributed to mixing with high-energy atomic orbitals and the coupling constant $\lambda_c$ is smaller than $\lambda_v$.\cite{Echeverry2016}

When the monolayer TMDC is placed on the ferromagnetic substrate, the magnetic exchange potential penetrates into the atomic layer, and
it leads to a Zeeman-type spin splitting depending on the magnetic moment in the substrate\cite{Hauser1969,White1985}.
When the ferromagnetic substrate is homogeneous, the induced potential is also homogeneous, and it is given by
\begin{align}
H_m=-\boldsymbol{M}\cdot\boldsymbol{s},\label{eq_exchange_potential}
\end{align}
where the exchange field $M_\mu$ is proportional to the magnetic moment in the ferromagnetic material.
The proximity Zeeman effect was reported in a non-magnetic two-dimensional material on ferromagnetic substrate e.g. EuO or EuS\cite{White1985,Swartz2012,Yang2013,Wei2016}.
A recent numerical calculation has shown the possibility to induce a large exchange potential $\sim40$meV in monolayer MoTe$_2$ on EuO\cite{Qi2015}.
Generally, the induced spin-dependent potential is smaller than $\Delta\sim O(1)$ eV and $\lambda_v\sim O(100)$meV, which characterize the insulating gap and the spin splitting, respectively.

\section{Anomalous Hall effect induced by out-of-plane magnetization}
\begin{figure}[htbp]
\begin{center}
 \includegraphics[width=70mm]{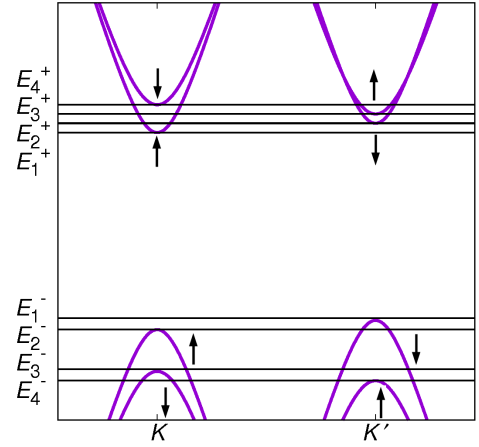}
\caption{Typical energy dispersion of TMDC on a ferromagnetic substrate. 
The band edges of the valence band and conduction band are represented by $E^-_j$ and $E^+_j$, respectively. Arrows indicate spins.
 }\label{Band_with_magnet}
\end{center}
\end{figure}

First, we consider the anomalous Hall effect induced by out-of-plane magnetization, $H_m=-M_zs_z$.
The exchange potential leads to the energy shift of electric states depending on its spin as,
\begin{align}
E^\pm_{s_z,\tau,k}=&\left(\frac{\lambda_-\tau}{2}-M_z\right)s_z\pm\frac{1}{2}\sqrt{(\Delta-\lambda_+\tau s_z)^2+4v^2k^2},
\end{align}
where $k=\sqrt{k_x^2+k_y^2}$ and $\lambda_\pm=\lambda_v\pm\lambda_c$.
We show the schematic picture of the band structure for a monolayer TMDC with the exchange potential in Fig.\ \ref{Band_with_magnet}.
The electronic conduction and valence band are split into four edges by $\lambda_\alpha$ and $M_z$,
\begin{align}
\begin{aligned}
E_1^+=&\frac{\Delta}{2}-\lambda_c-M_z\\
E_2^+=&\frac{\Delta}{2}-\lambda_c+M_z\\
E_3^+=&\frac{\Delta}{2}+\lambda_c-M_z\\
E_4^+=&\frac{\Delta}{2}+\lambda_c+M_z,
\end{aligned}
\end{align}
and
\begin{align}
\begin{aligned}
E_1^-=&-\frac{\Delta}{2}+\lambda_v+M_z\\
E_2^-=&-\frac{\Delta}{2}+\lambda_v-M_z\\
E_3^-=&-\frac{\Delta}{2}-\lambda_v+M_z\\
E_4^-=&-\frac{\Delta}{2}-\lambda_v-M_z,
\end{aligned}
\end{align}
respectively.

The Hall conductivity is a summation of the Berry phase over the occupied states as,
\begin{align}
\sigma_{xy}=&\frac{e^2}{2\pi h}\sum_{s_z,\tau}[\theta_{s_z,\tau}^+(E_F)+\theta_{s_z,\tau}^-(E_F)],
\end{align}
and
\begin{align}
\theta^\pm_{s_z,\tau}(E_F)=
\int d^2\boldsymbol{k}\;n_F(E^\pm_{s_z,\tau,k})\Omega^\pm_{s_z,\tau}(\boldsymbol{k}),\label{eq_Berry_phase}
\end{align}
where $\Omega^\pm_{\alpha}(\boldsymbol{k})=[\nabla_{\boldsymbol{k}}\times\langle u^\pm_\alpha,\boldsymbol{k}|i{\nabla}_{\boldsymbol{k}}|u^\pm_\alpha,\boldsymbol{k}\rangle]_z$ is the Berry curvature of the eigenstate $|u^\pm_\alpha,\boldsymbol{k}\rangle$ in the conduction band ($+$) and valence band ($-$), 
and $n_F(E)$ is the Fermi distribution function.
The out-of-plane exchange field merely shifts the energy band for each spin, and thus the Berry connection for $(s_z,\tau,\boldsymbol{k})$ is  equivalent to that in $M_z=0$\cite{Xiao2012}.
The Hall conductivity summed over the conduction and valence bands in each spin / valley sector can be explicitly written as,
\begin{align}
&\sigma_{xy}^{s_z,\tau}(E_F)=\frac{e^2}{2h}\tau\times\begin{cases}
\displaystyle{\frac{\Delta E}{|E_F-E_c|}}&(\Delta E<|E_F-E_c|)\\
1&(|E_F-E_c|\leq \Delta E)
\end{cases}\label{eq_Hall_conductivity}
\end{align}
where $E_c=(\lambda_-\tau/2-M_z)s_z$ is the center of gap in $(s_z,\tau)$ and $2\Delta E=\Delta-\lambda_+\tau$ is the gap energy.
In what follows, we consider two different cases,
molybdenum Mo$X_2$ and tungsten W$X_2$ ($X$=S, Se, Te),
which exhibit different qualitative features in the conduction band.
\subsection{Molybdenum dichalcogenide $\lambda_c\neq0$}
\begin{figure}[htbp]
\begin{center}
 \includegraphics[width=75mm]{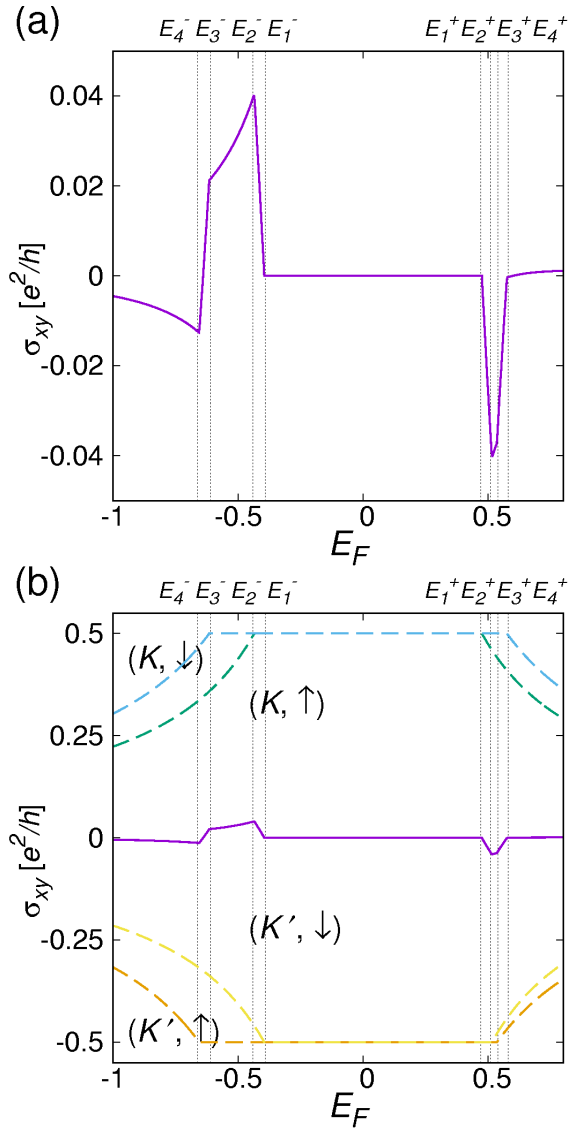}
\caption{(a) Hall conductivity of monolayer MoTe$_2$ with $M=20$meV plotted against the Fermi energy. (b) Contributions from each valley and spin sector.
 }\label{Hall_conductivity_0}
\end{center}
\end{figure}
We plot the net Hall conductivity $\sigma_{xy}(E_F)$ and the component parts $\sigma^{s_z,\tau}_{xy}(E_F)$ in Fig.\ \ref{Hall_conductivity_0}(a) and (b), respectively, where we assumed $M_z=20$meV and the material parameters of MoTe$_2$ in Table \ref{Hall_coefficient}.
Here, the exchange potential $M_z=20$meV is in the experimentally feasible range in the magnetic proximity effect simulated by a first principle calculation\cite{Qi2015}.
In the intrinsic TMDC, the Hall conductivity of $(K,s_z)$ exactly cancels with that of $(K',-s_z)$ because of time-reversal symmetry.
The exchange potential $M_z$ breaks the balance between them by shifting the band energies as shown in Fig.\ \ref{Band_with_magnet}.
In Fig.\ \ref{Band_with_magnet}(b), we actually see that the Hall conductivity curves of $(K,s_z)$ and $(K',-s_z)$ horizontally slide in the opposite directions, resulting in a finite net Hall conductivity.
The non-zero $\sigma_{xy}$ is observed only inside the conduction and valence band, i.e., finite doping, while the value inside the gap is quantized and never changes from zero.

The induced Hall conductivity is well approximated by the lowest order in the exchange potential $M_z$, under the realistic experimental condition of $M_z\ll \lambda_v$ and $M_z<\lambda_c$ for molybdenum compounds.
The approximate expressions at the band edges are given by
\begin{align}
\begin{aligned}
\sigma_{xy}(E_2^+)\simeq&-\frac{e^2}{h}\frac{2}{\Delta-\lambda_+}M_z\\
\sigma_{xy}(E_2^-)\simeq&\frac{e^2}{h}\frac{2}{\Delta-\lambda_+}M_z\\
\sigma_{xy}(E_3^-)\simeq&\frac{e^2}{h}\frac{2(\Delta-\lambda_+)}{(\Delta+3\lambda_+)^2}M_z\\
\sigma_{xy}(E_4^-)\simeq&\frac{e^2}{h}\left(\frac{2(\Delta-\lambda_+)}{(\Delta+3\lambda_+)^2}
-\frac{2}{\Delta+\lambda_+}\right)M_z.
\end{aligned}\label{Eq_Hall_conductivity_Mz}
\end{align}
\begin{table}[htb]
\caption{\label{tab:fonts}Band parameters for TMDCs used in the present calculation [\onlinecite{Kang2013},\onlinecite{Kormanyos2015},\onlinecite{Echeverry2016}], and the coefficients of $M_z$ linear term in the Hall conductivity, Eq.\ (\ref{Eq_Hall_conductivity_Mz}), in units of $(e^2/h)/$[eV].
}
\begin{center}
\begin{tabular}{c c c c c c c}
\hline
\hline
&MoS$_2$&MoSe$_2$&MoTe$_2$&WS$_2$&WSe$_2$&(WTe$_2$)\\ \hline
$\Delta$[eV]\cite{Kang2013,Kormanyos2015}&1.665&1.425&1.05&1.765&1.485&0.995\\ 
$\lambda_c$[eV]\cite{Echeverry2016}&0.008&0.018&0.029&0.001&0.001&-\\
$\lambda_v$[eV]\cite{Kang2013,Kormanyos2015}&0.075&0.095&0.11&0.215&0.235&0.245\\
$v$[eV$\cdot$\AA]\cite{Kormanyos2015}&2.76&2.53&2.33&3.34&3.17&3.04\\
$\sigma_{xy}(E_2^+)/M_z$&-1.26&-1.50&-2.13&-0.32&-0.46&-1.05\\  
$\sigma_{xy}(E_2^-)/M_z$&1.26&1.50&2.13&1.29&1.60&2.67\\ 
$\sigma_{xy}(E_3^-)/M_z$&0.89&0.91&0.99&0.53&0.52&0.50\\ 
$\sigma_{xy}(E_4^-)/M_z$&-0.26&-0.41&-0.74&-0.47&-0.64&-1.11\\ \hline \hline
\end{tabular}\label{Hall_coefficient}
\end{center}
\end{table}
Table\ \ref{Hall_coefficient} lists the coefficients of $M_z$-linear term in the Hall conductivity for several TMDCs,
where we use the material parameters in Ref.\ \onlinecite{Kang2013,Kormanyos2015,Echeverry2016}.
In any compounds, the Hall conductivity peaks at $E_F=E_2^\pm$ and $E_4^-$, and changes its sign between $E_3^-$ and $E_4^-$.
These characteristic behaviors can be observed within experimentally feasible carrier doping.
The carrier density corresponding to the characteristic band energies can be written in the lowest order of $M_z$ as,
\begin{align}
\begin{aligned}
\rho(E_2^+)\simeq&\frac{\Delta-\lambda_+}{\pi v^2}M_z,\\
\rho(E_2^-)\simeq&-\frac{\Delta-\lambda_+}{\pi v^2}M_z,\\
\rho(E_3^-)\simeq&-\frac{\Delta+\lambda_+}{\pi v^2}\left(\lambda_+-M_z\right),\\
\rho(E_4^-)\simeq&-\frac{1}{\pi v^2}\left((\Delta+\lambda_+)\lambda_++2(\Delta+3\lambda_+)M_z\right).
\end{aligned}
\end{align}
The required carrier density for the furthest energy $E_4^-$ at $M_z=0$ and $E_2^-$ at $M_z=20$meV is listed in Table\ \ref{Charge_density}.
\begin{table}[htb]
\caption{\label{tab:fonts}The carrier density for $E_F=E_2^-$ at $M_z=20$meV and $E_F=E_4^-$ at $M_z=0$ in units of $10^{14}$[cm$^{-2}$].}
\begin{center}
\begin{tabular}{c c c c c c c }
\hline \hline
&MoS$_2$&MoSe$_2$&MoTe$_2$&WS$_2$&WSe$_2$&WTe$_2$\\ \hline 
$\rho_0(E_2^-)$&-0.13&-0.13&-0.11&-0.08&-0.08&-0.05\\
$\rho_0(E_4^-)$&-0.55&-0.71&-0.74&-1.21&-1.28&-1.05\\  \hline\hline
\end{tabular}\label{Charge_density}
\end{center}
\end{table}

\subsection{Tungsten dichalcogenide $\lambda_c\simeq0$}
\begin{figure}[htbp]
\begin{center}
 \includegraphics[width=70mm]{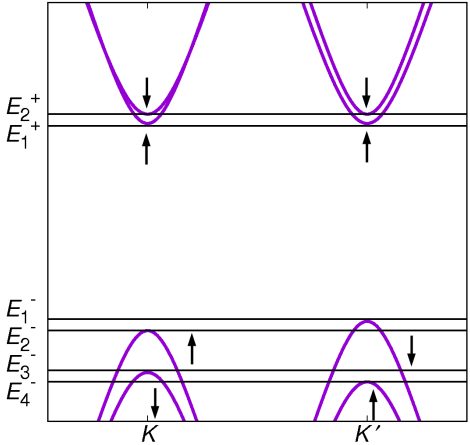}
\caption{Typical energy dispersion of W$X_2$ on a ferromagnetic substrate. Arrows indicate spins.
 }\label{Band_with_magnet_W}
\end{center}
\end{figure}
The Hall conductivity in the conduction band of tungsten dichalcogenides W$X_2$ is qualitatively different from the case in Mo$X_2$. 
The spin-splitting in the valence band, $\lambda_v$,
is much greater in W$X_2$ than in Mo$X_2$.
On the contrary, the conduction band splitting $\lambda_c$ is highly suppressed in 
W$X_2$ when the Fermi energy is crossing the conduction band.
\cite{Echeverry2016}.
The theoretical estimation of the spin splitting is  given in Table\ \ref{Hall_coefficient}, 
where $\lambda_c$  is negligibly small for W$X_2$.
Therefore, the band bottom in the conduction band is nearly spin-degenerate at $M_z=0$, and it is split into
\begin{align}
E_1^+ \simeq \frac{\Delta}{2}-M_z,\;\;E_2^+ \simeq \frac{\Delta}{2}+M_z,
\end{align}
as shown in Fig.\ \ref{Band_with_magnet_W}.

\begin{figure}[htbp]
\begin{center}
 \includegraphics[width=75mm]{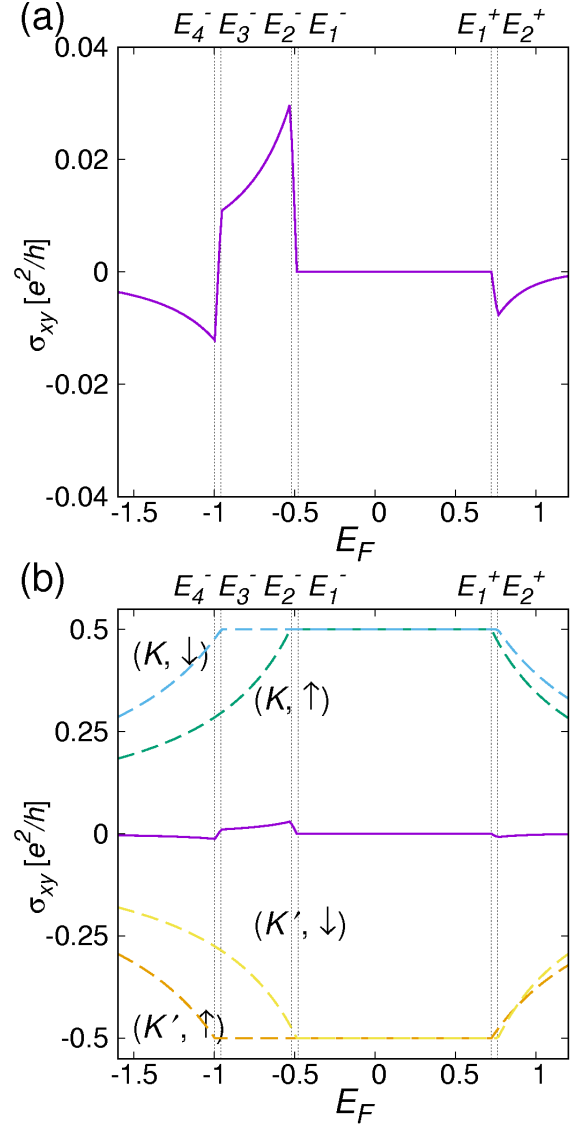}
\caption{(a) Hall conductivity of monolayer WSe$_2$ with $M=20$meV plotted against the Fermi energy. (b) Contributions from each valley and spin sector.
 }\label{Hall_conductivity_WSe2_0}
\end{center}
\end{figure}

In Fig.\ \ref{Hall_conductivity_WSe2_0}, we give the numerical result of Hall conductivity in WSe$_2$ with $M_z=20$meV. The behavior in the valence band is qualitatively similar to MoTe$_2$ in Fig.\ \ref{Hall_conductivity_0}.
On the other hand,  a non-zero Hall conductivity is still remaining in the conduction band, even though the band edges are nearly valley degenerate as shown in Fig.\ \ref{Hall_conductivity_WSe2_0}.
This is because the conduction bands of $K$ and $K'$ have the different band masses, and thus the different energy dependences of the Hall conductivity.

The Hall conductivity and charge density at the conduction band edge $E_2^+$ are approximately expressed as 
\begin{align}
\sigma_{xy}(E_2^+)\simeq&-\frac{e^2}{h}\frac{4\lambda_v}{\Delta^2-\lambda_v^2}M_z,
\end{align}
and
\begin{align}
\rho(E_2^+)\simeq&\frac{\Delta+\lambda_v}{\pi v^2}M_z,
\end{align}
respectively. The values estimated for several materials are listed in Table\ \ref{Hall_coefficient} and \ref{Charge_density}. Note that 2H structure is not the most stable phase in tungsten ditelluride WTe$_2$, and thus the fabrication of 2H-monolayer of WTe$_2$ needs the structural transition between 1T' to 2H.\cite{Huang2016}

\section{Effect of tilting of magnetization}
\begin{figure}[htbp]
\begin{center}
 \includegraphics[width=75mm]{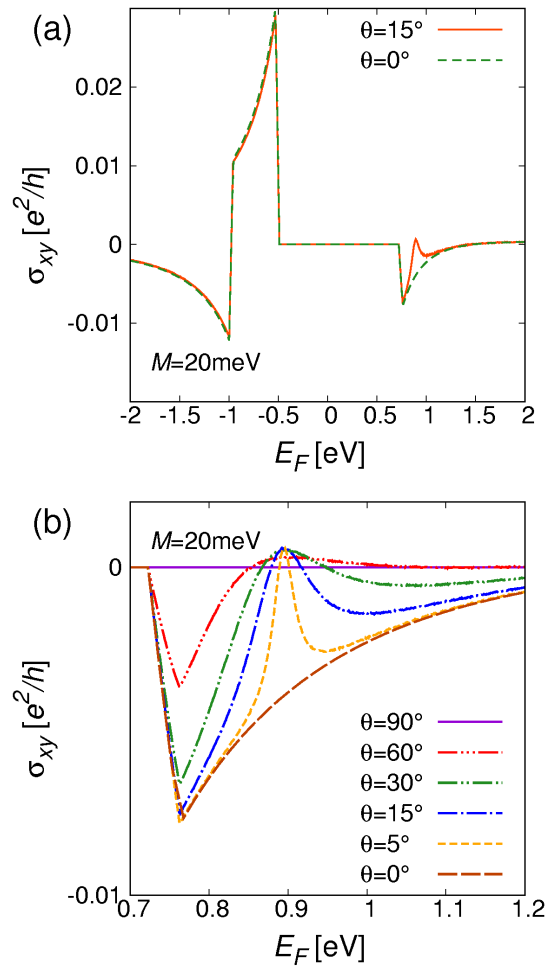}
\caption{(a) Hall conductivity against $E_F$ in monolayer WSe$_2$ with the exchange potential $M=$20meV with the tilting angle $\theta=0^\circ$ and $15^\circ$. (b) Similar plots near the conduction band bottom for $0^\circ<\theta<90^\circ$.
 }\label{Hall_conductivity_tilt}
\end{center}
\end{figure}
Next, we consider the Hall effect induced by a tilted magnetic moment to the atomic layer, where the magnetic exchange potential in Eq.\ (\ref{eq_exchange_potential}) couples to both the out-of-plane spin and in-plane spin components.
This effect is particularly important in TMDCs with small spin splitting in the conduction band, e.g., WX$_2$,
so in the following discussion we assume $\lambda_c=0$ for simplicity.
In Fig.\ \ref{Hall_conductivity_tilt}(a), we show the numerically calculated Hall conductivity for WSe$_2$ in $M=|\boldsymbol{M}|=20$ meV with tilting angle $\theta=0^\circ$ and $\theta=15^\circ$.
In the presence of the in-plane field, we observe a sharp peak slightly above the edge of the conduction band, at which $\sigma_{xy}$ reaches nearly zero.
There is no remarkable change in the hole side.
Fig.\ \ref{Hall_conductivity_tilt}(b) presents the plots around the conduction band bottom 
for different angles.
In increasing $\theta$, the peak width is gradually broadened, and at the same time the overall amplitude of the Hall conductivity is reduced in accordance with the decrease of the out-of-plane field $M_z=M\cos\theta$.
At $\theta=90^\circ$, the Hall conductivity is zero everywhere.
\begin{figure}[htbp]
\begin{center}
 \includegraphics[width=75mm]{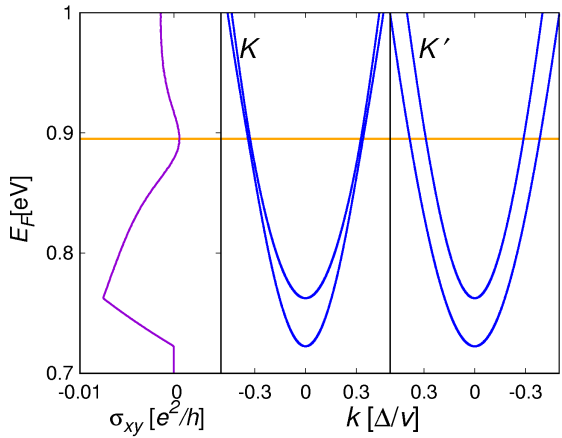}
\caption{
(Left) Hall conductivity near the conduction band bottom in WSe$_2$ monolayer with the exchange potential 20meV tilted by $\theta=15^\circ$. (Middle, Right) The band structures for $K$, $K'$, respectively, in the corresponding energy region.
 }\label{Hall_conductivity_3}
\end{center}
\end{figure}

In Fig.\ \ref{Hall_conductivity_3}, we compare the energy band and the Hall conductivity in the electron side, at $M=20$ meV and $\theta=15^\circ$.
In the band structure, we notice that the heavy band (spin down) and light band (spin up)
intersect at $K$ valley, and the position of the Hall conductivity dip actually coincides with the crossing point.
The band intersection appears only in $K$ valley because there $M_z$ shifts the heavy electron band upward and the light electron band downward.
In $K'$, the movement is opposite and no intersection occurs.

\begin{figure*}[htbp]
\begin{center}
 \includegraphics[width=160mm]{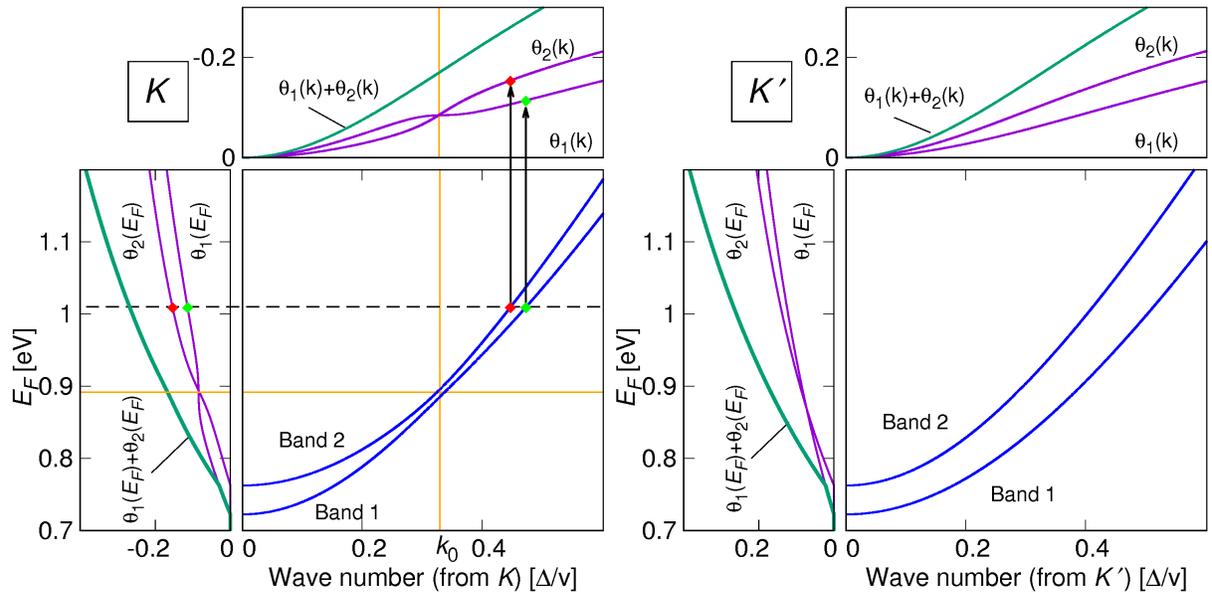}
\caption{Band structure and the Berry phase in $K$ and $K'$ valleys calculated for WSe$_2$ monolayer with $M=$20meV and $\theta=15^\circ$. The Berry phase is shown as a function of the wave vector in the top panel, and as a function of the Fermi energy in the left panel.
 }\label{Hall_conductivity_K}
\end{center}
\end{figure*}

To further explain the origin of the Hall conductivity dip, we plot in Fig.\ \ref{Hall_conductivity_K} the detailed band structure and the Hall conductivity components for different bands.
At $K$, the heavy band and light band are slightly anti-crossing due to the hybridization by the in-plane field $M_\parallel=M\sin\theta$.
At $K'$, on the other hand, the two bands remain isolated and the effect of $M_\parallel$ is minimal.
In each valley, we label the upper and lower branches as band 2 and 1, respectively.
The top figure in each panel plots $\theta_j(k)$, or the summation of the Berry curvature of band $j$ inside the Fermi circle with the radius $k$.
At $K$, there is a rapid interchange between $\theta_1$ and $\theta_2$ at the anti-crossing energy,
where $\theta_1$ switches from the Berry phase of the up-spin band to that of the down-spin band, and $\theta_2$ moves in the opposite direction.
The Berry phase $(\theta_1(k),\theta_2(k))$ can be written as a linear transformation of that at $M_\parallel=0$, denoted by $(\theta^{(0)}_\uparrow(k),\theta^{(0)}_\downarrow(k))$, as
\begin{align}
\begin{pmatrix}
\theta_{1}(k)\\
\theta_{2}(k)
\end{pmatrix}
=\frac{1}{2}\begin{pmatrix}
1+\frac{\Delta E^{(0)}_{k}}{\Delta E_{k}}&1-\frac{\Delta E^{(0)}_{k}}{\Delta E_{k}}\\
1-\frac{\Delta E^{(0)}_{k}}{\Delta E_{k}}&1+\frac{\Delta E^{(0)}_{k}}{\Delta E_{k}}
\end{pmatrix}
\begin{pmatrix}
\theta^{(0)}_{\uparrow}(k)\\
\theta^{(0)}_{\downarrow}(k)
\end{pmatrix}.\label{Berry_Phase_tilt}
\end{align}
Here, $\Delta E^{(0)}_k$ and $\Delta E_k$ are the energy differences between band 1 and 2 at $k$ for $M_\parallel =0$ and $M_\parallel\neq 0$, respectively,
which are defined by 
\begin{align}
&\Delta E_{k}^{(0)}= E^+_{\uparrow,\tau,k} - E^+_{\downarrow,\tau,k},\label{eq_Energy_diff_Mz}\\
&\Delta E_{k}= \sqrt{\left[\Delta E_{k}^{(0)}\right]^2+4{M_\parallel}^2}.\label{eq_Energy_diff_Mz+Mx}
\end{align}
It is straight forward to check
\begin{align}
\theta_1(k)+\theta_2(k)=\theta^{(0)}_\uparrow(k)+\theta^{(0)}_\downarrow(k).\label{eq_total_Berry_phase}
\end{align}
In Fig.\ \ref{Hall_conductivity_K}, we see that $\theta_1(k)+\theta_2(k)$ shows no special feature at the anti-crossing point, naturally because the sum does not depend on $M_\parallel$ according to Eq. (\ref{eq_total_Berry_phase}).
However, the Hall conductivity is defined by the summation of the Berry phase at the same Fermi energy, but not the same $k$.
In the left figures of Fig. \ref{Hall_conductivity_K}, we plot $\theta_j(E_F)$ by translating $k$ to $E_F$ using the energy dispersion.
Now we see that the summation $\theta_1(E_F)+\theta_2(E_F)$ slightly sinks at the anti-crossing point of $K$ valley.
This is because $E_F$ corresponds to the different $k$'s in the two bands (red and green points in Fig.\ \ref{Hall_conductivity_K}, and this breaks the cancellation of $M_\parallel$ dependent term in 
Eq.\ (\ref{eq_total_Berry_phase}).
While the dip in Fig.\ \ref{Hall_conductivity_K} looks tiny, the back ground part is mostly cancelled in the summation with $K'$, and we are left with a prominent dip observed in Fig.\ \ref{Hall_conductivity_3}.

\begin{figure}[htbp]
\begin{center}
 \includegraphics[width=75mm]{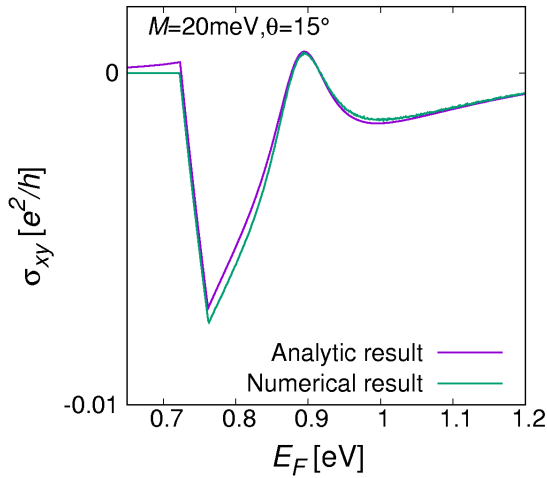}
\caption{Analytic and numerical plots of the Hall conductivity around the conduction band bottom in WSe$_2$ monolayer with $M=$20meV and $\theta=15^\circ$.
 }\label{Hall_conductivity_Lorentzian}
\end{center}
\end{figure}

Finally, we derive an analytic formulation of the dip structure in the Hall conductivity
by expanding Eq.\ (\ref{Berry_Phase_tilt}) at the band crossing point of $K$ valley. 
We define the momentum crossing point (i.e., $\Delta E^{(0)}_{k} = 0$) as $k_0$, and the corresponding energy as 
$E_0 \equiv E^+_{\uparrow,+,k_0} = E^+_{\downarrow,+,k_0}$.
In Fig.\ \ref{Hall_conductivity_K}, $k_0$ and $E_0$ are indicated by the vertical and horizontal solid lines, respectively.
The change of the Hall conductivity by $M_\parallel$ is well approximated by
a Lorentzian function in the Fermi energy as
\begin{align}
&\Delta \sigma_{xy} \approx \frac{e^2}{2\pi h}
\left[	\theta^{(0)}_{\uparrow}(k_0)-\theta^{(0)}_{\downarrow}(k_0)\right]
\frac{\alpha}{1+(E_F-E_0)^2\alpha^2/M_\parallel^2},
\label{Lorenzian}
\end{align}
where
\begin{align}
&\alpha = \frac{v_\uparrow + v_\downarrow}{v_\uparrow - v_\downarrow},
\end{align}
and $v_\uparrow$ and $v_\downarrow$ are the band velocities of $E^+_{\uparrow,+,k}$ and $E^+_{\downarrow,+,k}$
at $k=k_0$, respectively.
After some algebra, we have,
\begin{align}
&\theta^{(0)}_{\uparrow}(k_0)-\theta^{(0)}_{\downarrow}(k_0)
= \frac{8\pi M_z\Delta(\lambda - 2M_z)(\lambda - M_z)}{(\lambda\Delta)^2-(\lambda - 2M_z)^4},
\nonumber\\
& \alpha = \frac{(\lambda - 2M_z)^2}{\lambda\Delta},
\end{align}
where $\lambda= \lambda_v$.
In Fig.\ \ref{Hall_conductivity_Lorentzian},
we plot the total Hall conductivity calculated by adding Eq.\ (\ref{Lorenzian}) to Eq.\ (\ref{eq_Hall_conductivity}),
which nicely agrees with the numerical result.
When $M_z \ll \lambda_v$,  Eq.\ (\ref{Lorenzian}) is even reduced to
\begin{align}
&\Delta \sigma_{xy} \approx \frac{e^2}{h}
\frac{4\lambda M_z}{\Delta^2-\lambda^2}
\frac{1}{1+(E_F-E_0)^2\lambda^2/(M_\parallel\Delta)^2}.
\end{align}
The width of the Lorentzian is given by $M_\parallel \Delta /\lambda$,
and it is broadened linearly in increasing the in-plane magnetization $M_\parallel$.
The height, $(e^2/h)4\lambda M_z/(\Delta^2-\lambda^2)$,
coincides with $\sim-\sigma_{xy}(E_2^+)$ in Eq. (\ref{Eq_Hall_conductivity_Mz}), and this is why the total Hall conductivity nearly vanishes at the peak center.
The band crossing and the sharp behavior in Hall conductivity
also occur in molybdenum compound, but it appears at higher Fermi energy due to the intrinsic spin-orbit splitting 
$\lambda_c$.

\section{Conclusion}
In conclusion, we consider the anomalous Hall effect in monolayer 2H-TMDCs under the external exchange field, 
and propose that the Hall conductivity can be induced by the magnetic proximity effect.
The induced Hall conductivity is found to be of the order of $(e^2/h)(M_z/\Delta)$, and it should be observable 
in the realistic situation with $M_z \sim$  a few 10 meV and $\Delta \sim 1$ eV.
We also found that the tilt of the magnetization induces a sharp peak structure in the Hall conductivity 
in the electron side as a function of the Fermi energy, 
and explained it in terms of the Berry phase mixing by the in-plane field.
The present work 
provides a simple and independent method to 
detect the intrinsic Berry phase of the 2H-TMDC monolayers by usual Hall measurement.
The proximity Hall effect should generally occur in other 2D materials with the spin-orbit coupling, and we expect that this would be used as a general approach to probe the intrinsic Berry phase.
The detailed study on the proximity Hall effect in other 2D materials is left for future work.

\bibliography{TMD}
\end{document}